\documentclass[a4paper,11pt]{article}
\usepackage{latexsym}	
\usepackage{amsmath}
\usepackage{amsthm}

\usepackage{graphicx}
\def\ba{\begin{eqnarray}}
\def\ea{\end{eqnarray}}

\def\ba{\begin{eqnarray}}
\def\ea{\end{eqnarray}}

\def\lb{\label}
\def\be{\begin{equation}}
\def\ee{\end{equation}}

\theoremstyle{plain}

\begin{document}

\title{Peculiarities for scalar fields in GR domain walls}
\author{Andrés Felipe Tovar $^1$\thanks{
andresfelipe630@gmail.com.} \and Osvaldo Pablo Santillán$^2$\thanks{firenzecita@hotmail.com and osantil@dm.uba.ar.}}
\date{%
    $^1$Departamento de F\'isica, FCEyN, Universidad de Buenos Aires, Argentina\\%
    $^2$Instituto de Matem\'atica Luis Santal\'o (IMAS), UBA CONICET, Buenos Aires, Argentina\\[2ex]%
    }
\maketitle
\begin{abstract}
In the present letter, the solutions of the Klein Gordon equation in
the space time of an infinite domain wall is studied. It is shown
that the horizon is an special region, where initial conditions (scattering conditions) at asymptotic past
for the field can not be defined. This is in harmony with the fact that this region
can not be crossed, a possibility that was suggested in \cite{lanosa}. If these results are correct, then  it makes no sense to impose boundary conditions at this location.

 \end{abstract}
 
The present work deals with an infinite vacuum domain wall in GR, that is, the solution of the Einstein equations for matter distributed along a plane, denoted in the following as $yz$. The energy momentum tensor of such configuration is independent
of the $x$ coordinate and is locally given by  $T_{\mu\nu}=\mu (1,0,1,1)\delta(x)$ \cite{vilenkin1}. The corresponding gravitational field have been found in \cite{vilenkin2} and it is described locally by the metric 
\be\lb{domwall}
g=e^{-\mu |x|} (-dt^2+dx^2)+e^{\mu(t-|x|)}(dy^2+dz^2).
\ee
Here $\mu$ is the surface mass density on the object, whose location is at $x=0$. The Newtonian limit shows that the gravitational field of the wall is indeed repulsive \cite{vilenkin2}.
The fact that the metric decays exponentially 
at infinite is non physical, and it is showing that these coordinates do not cover the whole space. However, coordinates that go beyong the region $x\to \pm \infty$ can be found separately for both the left
$x<0$ or the right hand $x>0$. For positive values of $x$, this coordinate should be replaced by
$$
\Xi=\frac{2}{\mu}(1-e^{-\frac{\mu x}{2}}).
$$
The new coordinate $\Xi$ takes values in the interval $[0, 1)$ when $x$ takes all the positive real values. An analogous coordinate can be constructed for negative $x$ values.
By further extending the range of $\Xi$ to the interval $(-\infty,\infty)$, the metric becomes
$$
g=-\bigg(1-\frac{\mu |\Xi|}{2}\bigg)^2 dt^2+d\Xi^2+e^{\mu t}\bigg(1-\frac{\mu |\Xi|}{2}\bigg)^2(dy^2+dz^2),
$$
This metric goes beyond the region $x\to \pm \infty$, and is an extension of (\ref{domwall}) to a larger region.
There are two horizons  at $\Xi=\pm 2/\mu$. Any of those horizons corresponds to the apparent asymptotic region to $x\to \pm \infty$, which is in fact at finite
distance from the wall at $x=0$  \cite{vilenkin1}-\cite{vilenkoso2}.

Although the last is a valid extension of the metric, it was shown in \cite{lanosa} that there are further issues related to its global behaviour.
In particular, it was suggested that the horizon is not reachable neither from inside or outside. In order to support the points of view of \cite{lanosa}, the behaviour
of an scalar field in the inner part of the geometry will be considered, and it will be shown to be pathological when reaching the horizon region. 

Consider first the region on the right of the wall, inside the horizon. If the results of \cite{lanosa} are correct, this region should be enough for the present purposes, as the scalar field $\phi$
will not cross neither of the two horizons. The region in consideration then corresponds to $x>0$.   From the metric tensor \eqref{domwall} it follows that
\[
g_{\mu\nu}=e^{-\mu x}\begin{pmatrix}-1\\
    & 1\\
    &  & e^{\mu t}\\
    &  &  & e^{\mu t}
   \end{pmatrix}\quad g^{\mu\nu}=e^{\mu x}\begin{pmatrix}-1\\
    & 1\\
    &  & e^{-\mu t}\\
    &  &  & e^{-\mu t}
   \end{pmatrix},\quad\det g=-e^{-4\mu x}e^{2\mu t}.
   \]
On the other hand the equations of motion for a free scalar field $\phi$ in a curved geometry is given by
$$
    \left[\frac{1}{\sqrt{-g}}\partial_{\mu}\left(\sqrt{-g}g^{\mu\nu}\partial_{\nu}\right)+m^{2}\right]\phi\left(x\right)=0,
    $$
where $m$ denotes its mass. For the diagonal metric given above, the laplacian terms in the Klein Gordon equation takes the form
$$
\frac{1}{\sqrt{-g}}\partial_{\mu}\left(\sqrt{-g}g^{\mu\nu}\partial_{\nu}\right)=\frac{1}{\sqrt{-g}}\partial_{\mu}\left[\sqrt{-g}\left(g^{\mu0}\partial_{0}+g^{\mu1}\partial_{1}+g^{\mu2}\partial_{2}+g^{\mu3}\partial_{3}\right)\right]
    $$
    $$
    =e^{2\mu x}e^{-\mu t}\partial_{\mu}\left[e^{-2\mu x}e^{\mu t}\left(-e^{\mu x}\delta^{\mu0}\partial_{0}+e^{\mu x}\delta^{\mu1}\partial_{1}+e^{\mu\left(x-t\right)}\delta^{\mu2}\partial_{2}+e^{\mu\left(x-t\right)}\delta^{\mu3}\partial_{3}\right)\right]
    $$
    \be\lb{rial}
    =e^{2\mu x}e^{-\mu t}\partial_{0}\left(-e^{\mu\left(t-x\right)}\partial_{0}\right)+\partial_{1}\left(e^{\mu\left(t-x\right)}\partial_{1}\right)+\partial_{2}\left(e^{-\mu x}\partial_{2}\right)+\partial_{3}\left(e^{-\mu x}\partial_{3}\right).
\ee
In addition, due to the translational symmetry along the plane $yz$ the following variable separation may be postulated for the scalar field $\phi$
\begin{equation}
    \phi(t,x,y,z)=e^{i k_y y +i k_z z}T\left(t\right)X\left(x\right).
\end{equation}
The insertion of this into  the Klein Gordon equation, by taking into account \eqref{rial} and the identities
\begin{align*}
    -e^{2\mu x}e^{-\mu t}\partial_{0}\left(e^{\mu\left(t-x\right)}\partial_{0}\right)&=-e^{\mu x}\mu\partial_{0}-e^{\mu x}\partial_{0}^{2},\\
    e^{2\mu x}e^{-\mu t}\partial_{1}\left(e^{\mu\left(t-x\right)}\partial_{1}\right)&=-\mu e^{\mu x}\partial_{1}+e^{\mu x}\partial_{1}^{2},\\
    e^{2\mu x}e^{-\mu t}\partial_{2}\left(e^{-\mu x}\partial_{2}\right)&=e^{\mu\left(x-t\right)}\partial_{2}^{2},
\end{align*}
 results into the following differential equation
$$
-e^{\mu x}\bigg(\frac{d^{2}T(t)}{dt^{2}}\frac{1}{T}+\mu\frac{dT(t)}{dt^{2}}\frac{1}{T(t)}\bigg)+e^{\mu x}\bigg(\frac{d^{2}X(x)}{dx^{2}}\frac{1}{X(x)}
-\mu\frac{dX(x)}{dx}\frac{1}{X(x)}\bigg)
- e^{\mu(x-t)}(k_{y}^{2}+k_{z}^{2})+m^{2}=0.
$$
The standard trick of variable separation leads to the two equations
\[\begin{split}
    \left[\frac{d^{2}}{dt^{2}}+\mu\frac{d}{dt}+e^{-\mu t}\left(k_{y}^{2}+k_{z}^{2}\right)+\alpha\right]T\left(t\right)&=0,\\
    \left[\frac{d^{2}}{dx^{2}}-\mu\frac{d}{dx}+m^{2}e^{-\mu x}+\alpha\right]X\left(x\right)&=0,
\end{split}\]
with $\alpha$ an integration constant. These are linear differential equations that may be solved as follows.

 Consider the right hand side of the wall, that is the region $x>0$.
With the change of variables $v=e^{-\frac{\mu t}{2}}$ and the identity
\[
    \begin{split}
        \frac{d}{dt}\left(\frac{d}{dt}\right)=\frac{dv}{dt}\frac{d}{dv}\left(\frac{d}{dv}\frac{dv}{dt}\right)=\frac{dv}{dt}\left[\frac{d}{dv}\left(\frac{dv}{dt}\right)\frac{d}{dv}+\frac{dv}{dt}\frac{d^{2}}{dv^{2}}\right]\\
        =\left(\frac{\mu}{2}\right)^{2}v^{2}\frac{d^{2}}{dv^{2}}+\left(\frac{\mu}{2}\right)^{2}v\frac{d}{dv}
    \end{split}\]
the following temporal equation is obtained   
 $$
        \left[v^{2}\frac{d^{2}}{dv^{2}}-v\frac{d}{dv}+\left(\frac{2}{\mu}\right)^{2}v^{2}\left(k_{y}^{2}+k_{z}^{2}\right)+\left(\frac{2\sqrt{\alpha}}{\mu}\right)^{2}\right]T\left(v\right)=0.
        $$
 In analogous fashion the variable change $u=e^{-\frac{\mu x}{2}}$ leads to   
        $$
        \left[u^{2}\frac{d^{2}}{du^{2}}+3u\frac{d}{du}+\left(\frac{2m}{\mu}\right)^{2}u^{2}+\left(\frac{2\sqrt{\alpha}}{\mu}\right)^{2}\right]X\left(u\right)=0.
    $$   
The last two are of the form of the Bowman equations \cite{bowman}
    \[x^{2}\frac{d^{2}y}{dx^{2}}+\left(2p+1\right)x\frac{dy}{dx}+\left(\alpha^{2}x^{2r}+\beta^{2}\right)=0\]
whose solution is given by
\[ y\left(x\right)=x^{-p}\left[C_{1}J_{\frac{q}{r}}\left(\frac{\alpha}{r}x^{r}\right)+C_{2}N_{\frac{q}{r}}\left(\frac{\alpha}{r}x^{r}\right)\right],
\quad q\equiv\sqrt{p^{2}-\beta^{2}}.\]
Here the functions involved are Bessel functions when $\frac{q}{r}$ is real but are less studied when this parameter is imaginary \cite{dok}-\cite{rusos}.
In particular, their asymptotic behaviour of these solutions depends whether $q$ is real or imaginary. Even for the imaginary case, these functions will be loosely called Bessel functions in the following, even taking into account that some authors may prefer a different terminology.

In these terms, the more general solution for $\left(2.6\right)$ and $\left(2.7\right)$ is given by
$$
T(v)=v\bigg[C_{1} J_{{\scriptscriptstyle \sqrt{1-\frac{4\alpha}{\mu^{2}}}}}\bigg(\frac{{\scriptstyle 2\sqrt{k_{y}^{2}+k_{z}^{2}}}}{\mu}v\bigg)+ C_{2}N_{{\scriptscriptstyle \sqrt{1-\frac{4\alpha}{\mu^{2}}}}}\bigg(\frac{{\scriptstyle 2\sqrt{k_{y}^{2}+k_{z}^{2}}}}{\mu}v\bigg)\bigg].
$$
$$
X(u)=u^{-1}\bigg[C_{1}^{\prime}J_{{\scriptscriptstyle \sqrt{1-\frac{4\alpha}{\mu^{2}}}}}\bigg({\scriptstyle \frac{2m^{2}}{\mu}}u\bigg)+ C_{2}^{\prime}N_{{\scriptscriptstyle \sqrt{1-\frac{4\alpha}{\mu^{2}}}}}\bigg({\scriptstyle \frac{2m^{2}}{\mu}}u\bigg)\bigg].
$$
where $C_{1},C_{2},C_{1}^{\prime}$ and $C_{2}^{\prime}$ are normalizaton constants and $J_\nu(x)$, $N_\nu(x)$ are the above described Bessel functions. For real values of the order $\nu$ they are usually named Bessel functions of first and second species respectively. 
Their behavior at the origin is given by
\[\lim_{x\to0}J_{\nu}\left(x\right)=\frac{1}{\Gamma\left(\nu+1\right)}\left(\frac{x}{2}\right)^{\nu}
,\quad\lim_{x\to0}N_{\nu}\left(x\right)=-\frac{\Gamma\left(\nu\right)}{\pi}\left(\frac{2}{x}\right)^{\nu}
    \]
Here  $\nu=\sqrt{1-\frac{4\alpha}{\mu^{2}}}$. The horizon is located as $u\to 0$ and the last formulas imply that $X(u)$ is not divergent only if $\nu\geq 1$
and $C_{2}^{\prime}=0$. The first of these conditions implies that $\alpha\leq 0$. Therefore 
$$
X(u)=\frac{C_{1}^{\prime}}{u}J_{{\scriptscriptstyle \sqrt{1-\frac{4\alpha}{\mu^{2}}}}}\bigg({\scriptstyle \frac{2m^{2}}{\mu}}u\bigg)\longrightarrow 0,\qquad \text{when}\qquad  u\to 0.
$$
except if $\alpha=0$, in this case $X(u)$ tends to a constant.  If instead $\mu^2\geq 4\alpha$ but $\alpha$ is positive, then the inspection of the formulas given above for the near origin behavior show that $X(u)$ is divergent. In other words, there is no way to insure a finite value at the origin, except if $\alpha$ vanish.  In the opposite case $\mu^2\leq 4\alpha$ or equivalently, when $\nu$ is imaginary,
the solutions are known to posses a behavior near the origin $x\sim 0$ of the following  approximated form \cite{dok}-\cite{rusos}
$$
J_\nu(x)\sim \cos(\log \frac{x}{2}), \qquad N_\nu(x)\sim \sin(\log \frac{x}{2}).
$$
In other words, these basis of functions have a singularity near the origin, since they are not defined at this location. The function $X(u)$
is obtained by a product of these functions and a factor $u^{-1}$ and therefore, its value is not defined in the origin $u\sim 0$, as it is the product of a divergent factor with a highly oscillating function that may reach a zero or not.

The conclusion above is that $X(u)$ at the horizon can take values $0$, $\infty$ or not defined, except if the integration constant is $\alpha=0$. This already posses a difficulty for studying the scattering problem where a prescribed field value $\phi_0$ is defined at the horizon at the asymptotic past, as it seems that it can not be described in terms of the above eigenfuctions. It is interesting to study further the behavior of $T(v)$ at $v\to \infty$ which, as discussed above, corresponds to the earlier times $t\to -\infty$.
This requires the study of the Bessel functions for large arguments $x\to\infty$. The corresponding behavior, for real $\nu$, goes as follows
$$
J_{\nu}(x)\sim \sqrt{\frac{2}{\pi x}} \cos\bigg(x-\frac{\nu \pi}{2}-\frac{\pi}{4}\bigg),\qquad N_{\nu}(x)\sim \sqrt{\frac{2}{\pi x}} \sin\bigg(x-\frac{\nu \pi}{2}-\frac{\pi}{4}\bigg),
$$
Since the asymptotic past $t\to -\infty$ corresponds to $v\to \infty$ the last asymptotic formula and the definition given above of $T(v)$
implies that $T(v)\to\pm \infty$ at earlier times regardless the choice of the constants $C_1$ and $C_2$. Note that this asymptotic is a consequence not only of the last formula, but follows by taking into account that the definition of $T(v)$ includes a linear factor $v$. The conclusion is then that there is no way to make this quantity finite at 
the infinite past $t\to-\infty$. This is of course valid for $4\alpha<\mu^2$ since this corresponds to the case in consideration, for which $\nu=\sqrt{1-\frac{4\alpha}{\mu^2}}$ is real. In this situation the
scalar field
\begin{equation}
    \phi(t,x,y,z)=e^{i k_y y +i k_z z}T\left(t\right)X\left(x\right),
\end{equation}
will be divergent or will posses a $0.\infty$ indetermination at the asymptotic past $t\to -\infty$ and at the horizon position $x\to \infty$, for finite values of the coordinates  $yz$.
This is clear since $X(u)$, as stated above, is divergent or vanishes, except for $\alpha=0$, in which becomes a constant.

The remaining case is when $\nu=\sqrt{1-\frac{4\alpha}{\mu^2}}$ takes imaginary values, that is,  for values of the integration constant $\alpha$ such that
$4\alpha>\mu^2$. The asymptotic behavior of $T(v)$ in this case is not that well studied
\cite{dok}-\cite{rusos}. However, this function will be multiplied by a function $X(u)$ that has not defined values at the horizon position, thus
a $0.\infty$ indetermination, or a non determined behavior in general, is to be expected. This means that at this location
the field $\phi$ is not defined, and it may depend on the curve $x(t)$ chosen to take the limit. 

The argument given above suggest that the horizon is not a reachable point of the space time. The reader may argue this point of view, and think instead
that this behavior reflects that the coordinates $t$, $x$, $y$ and $z$ are ncomplete ones, which do not cover the full space time.
While this is a true statement, the results given above are in harmony with the analysis presented in \cite{lanosa}. In that reference the horizon is described
as a region inside a plane which runs away at light speed, thus a massive particle will not reach it. Furthermore, there are flat coordinates
such that the inner region between the wall and the horizon corresponds to positions $X_h<X$ with $X_h$ the horizon position. The asymptotic past is described in this reference
as $X_h\to \infty$. Clearly, it is not possible to describe the region $X_h<X$ in this limit, the description will be ill defined. Therefore, while the results given above are not complete proofs,
they are to be expected if the description of the reference \cite{lanosa} is indeed correct.

\section*{Acknowledgments}
O.P.S supported by CONICET, Argentina and  by the Grant PICT 2020-02181.

\end{document}